\documentclass[twocolumn,showpacs]{revtex4}
\usepackage[xdvi]{graphicx}%
\usepackage{dcolumn}
\usepackage{amsmath}
\makeatletter

\def\btt#1{\texttt{\@backslashchar#1}}%
\DeclareRobustCommand\bblash{\btt{\@backslashchar}}%
\makeatother

\newcommand{\bra}{\left\langle}
\newcommand{\ket}{\right\rangle}

\newcommand{\pder}[2]{\frac{\partial #1}{\partial  #2}}

\newcommand{\veck}{{\boldsymbol{k}}}
\newcommand{\vecr}{{\boldsymbol{r}}}
\newcommand{\vecv}{{\boldsymbol{v}}}
\newcommand{\vece}{{\boldsymbol{e}}}

\begin{document}
\title{Anomalous pressure in fluctuating shear flow}
\author{Hirofumi Wada and Shin-ichi Sasa}
\affiliation
{Department of Physics, University of Tokyo, Hongo, Tokyo 113-0033, Japan,\\
Department of Pure and Applied Sciences, University of Tokyo, 
Komaba, Tokyo 153-8902, Japan}

\date{\today}

\begin{abstract}
We investigate how the pressure in fluctuating shear flow depends on 
the shear rate $S$ and on the system size $L$ by studying fluctuating 
hydrodynamics under shear conditions. We derive anomalous forms 
of the pressure for two limiting values of the dimensionless parameter 
$\lambda=SL^2/\nu$, where $\nu$ is the kinetic viscosity. 
In the case $\lambda \ll 1$, the pressure is not an intensive quantity because
of the influence of the long-range spatial correlations of momentum 
fluctuations. 
In the other limit $\lambda \gg 1$, the long-range correlations are 
suppressed at large distances, and the pressure is intensive. 
In this case, however, there is the interesting effect that the
non-equilibrium correction to the pressure is proportional to $S^{3/2}$,
which  was previously obtained with the projection operator method 
[K. Kawasaki and J. D. Gunton, Phys. Rev.  {\bf A 8}, 2048, (1973)]. 
\end{abstract}

\maketitle

%%% intro %%%

It has been confirmed over a century of study that hydrodynamic equations 
describe macroscopic flow with high accuracy. However, a microscopic 
foundation of fluid mechanics has not yet been established except 
in the dilute limit gas. There seems to be no basis in microscopic physical 
laws on which to establish the validity of the local equilibrium assumption 
inherent in hydrodynamic equations. Indeed, it may be the case that any proper 
equation of state must be incorporated with non-equilibrium effects.

The anomalous form of equation of state under non-equilibrium conditions 
was first discussed  by Kawasaki and Gunton \cite{KG}.  They derived a 
non-analytic dependence of the pressure tensor on the shear rate for an 
uniformly sheared simple fluid.  
Subsequently, the same problem was studied by several authors in
detail \cite{KG, YK, Ernst, Lutsko}, and the non-analytic 
response due to mode-coupling effects has now become evident. 
Nevertheless, with regard to the normal stress, such predictions 
have not been fully confirmed 
by laboratory experiments or by numerical simulations \cite{Mar}.

In addition, quite recently, another type of anomalous non-equilibrium 
pressure has been reported by Aoki and Kusnezov \cite{Aoki}. They have 
studied numerically heat conduction problems of anharmonic oscillator 
models and have found that the nonequilibrium correction to the pressure 
is nearly proportional to the system size.  This finding is remarkable 
because it implies the absence of the intensivity of the pressure in 
this nonequilibrium system. As far as we know, there is no theoretical 
explanation for this result so far. 

In this Letter, we propose a unified understanding of the two anomalies 
for a specific simple case, uniform shear flow in a fluid.  
We establish a general criterion which can distinguish between 
non-intensive and non-analytic nature of the pressure.
We also give a brief comment on numerical experiments 
which disagree with the mode-coupling theory. 

The key idea of our study is to establish a possible relationship 
between the pressure anomalies and the long-range spatial correlation 
of momentum fluctuations.  It has been recognized that a lack of detailed 
balance is responsible for various distinctive features. One of them 
is the generic existence of long-range spatial correlations of 
fluctuations of conserved quantities \cite{LRC}.
For example, a correlation of momentum fluctuations becomes
spatially long-ranged under shear conditions \cite{Opp5,Dufty}.  

However, because the correlation function known at present is strongly 
divergent in a long wavelength limit, a certain appropriate length 
scale may be introduced as a cut off.  
Therefore we first need to know the momentum correlation function over 
the whole region of length scales.  
It may be suitable to study fluctuating hydrodynamics for this purpose. 

\paragraph*{Model:}

Let $v_i(\vecr,t)$ be a fluctuating velocity field in an incompressible fluid 
with a constant temperature $T$ that is far from the critical point.
The time evolution of $v_i(\vecr,t)$ is described by the 
continuity equation of the momentum,
\begin{equation}
\rho\partial_t{v_i}+\partial_j \Pi_{ij}  = 0,
\label{pcon}
\end{equation}
where the momentum flux tensor $\Pi$ is given by 
\begin{equation}
\Pi_{ij} = \rho v_i v_j+p\delta_{ij}-\nu \rho
(\partial_iv_j+\partial_jv_i) +s_{ij}.
\label{ns}
\end{equation}
Here, $\nu$ is the kinematic viscosity, $\rho$ is a constant density, and 
$s$ represents a Gaussian random stress tensor satisfying the
fluctuation-dissipation relation \cite{LL},
\begin{eqnarray}
\bra s_{ik}(\vecr,t) s_{lm}(\vecr',t')\ket
&=&  2\rho T \nu\left[(\delta_{il}\delta_{km}+\delta_{im}\delta_{kl})
\right. \nonumber \\
&-& \!\!\! \left.\frac{2}{3}\delta_{ik}\delta_{lm}\right]
\delta^3(\vecr-\vecr')\delta(t-t').
\label{ni}
\end{eqnarray}
Throughout this Letter, the Boltzmann constant is set to unity.
The auxiliary field $p$ in (\ref{ns}) can be replaced by use of the
incompressible condition $\partial_l v_l = 0$, which implies the relation
\begin{equation}
p=p_{\rm B}-\partial_l\partial_m s_{lm}-
\rho \Delta^{-1}(\partial_l v_m )(\partial_m v_l ),
\label{p}
\end{equation}
where $p_{\rm B}$ is a constant.

We consider the system in a three dimensional space, for which 
$-\infty < x,z  < \infty$ and $ -L/2 \le y \le L/2$.
The boundary conditions are imposed satisfying 
\begin{equation}
\vecv(x,L/2,z) = \vecv(x,-L/2,z)+SL \vece_x,
\end{equation}
which are chosen so as to make the analysis as simple as possible.
We study linear fluctuations around the average shear flow 
\begin{equation}
\bra \vecv(x,y,z) \ket=(Sy,0,0).
\end{equation}
Without loss of generality, we set $S \ge 0$.  

{}From (\ref{pcon}), we have $\partial_j \bra \Pi_{yj}\ket  = 0 $ for
the statistical steady state. 
Because this represents a force balance equation, the pressure in the 
$y$ direction is given by $\bra \Pi_{yy} \ket$. 
Then, using (\ref{ns}) and (\ref{p}), we write the pressure $p_y$ as
\begin{equation}
p_y = p_{\rm B}+\rho \bra v_y^2 \ket-\rho \Delta^{-1}
\bra (\partial_l v_m )(\partial_m v_l )\ket.
\label{pre}
\end{equation}
Now, the problem is to determine how $p_y$ depends on the 
shear rate $S$ and on the system size $L$.

The most general form of fluctuating hydrodynamics consists of a set of 
stochastic evolution equations for the energy, density and momenta \cite{LL}.  
In our model, the thermal diffusion constant and 
the sound damping constant are assumed to be much larger than the 
kinematic viscosity. This assumption is made for simplicity here,
otherwise the calculation of the stress tensor 
becomes more complicated \cite{Ernst, Lutsko}.

Now, we make a comment on the local equilibrium assumption that is involved in
hydrodynamic descriptions. 
Let $\ell$ be a microscopic length, such as a mean free path, and define a 
dimensionless parameter $\epsilon$ as
\begin{equation}
\epsilon = \frac{\ell^2 S }{\nu}.
\label{eqd}
\end{equation}
In order to investigate the physical meaning of this parameter, we first 
consider the case $\epsilon \to 0$. 
In this case, the local equilibrium assumption is valid 
because of the separation of scales.
This implies that the shear flow does not influence the thermodynamic 
properties of the system. 
However, 
if $\epsilon$ becomes of order unity, the molecular motion is violently 
disturbed by the shear. 
Although it may be interesting to study such systems, we cannot use 
fluctuating hydrodynamics for that aim. 
From this brief consideration of these two limiting cases, we regard the 
parameter $\epsilon$ as representing the extent of departure from the condition
of local equilibrium states. 
Because we seek the non-equilibrium correction to the pressure under shear 
within the framework of fluctuating hydrodynamics, we consider the case of 
a small but finite $\epsilon$.

As the final condition on our model, we assume $L \gg \ell$, so that the 
continuum description is valid. 
Note that our model does not include the parameter $\ell$ explicitly. 
It is included only as a cut-off length when we encounter an ultraviolet 
divergence. 

\paragraph*{Dimensional analysis :}

There are two independent dimensionless parameters in the system:
\begin{equation}
\lambda = \frac{L^2 S}{\nu}, \qquad
D =  \frac{T}{\rho \nu^2L}.
\end{equation}
Quite generally, we can write
\begin{equation}
p_y-p_y^{\rm eq}=\frac{T}{L^3}
f\left(\frac{L^2 S}{\nu}, \frac{T}{\rho \nu^2 L}\right),
\label{fscale}
\end{equation}
where $p_y^{\rm eq}$ is the equilibrium pressure.
We have assumed that $p_y-p_y^{\rm eq}$ does not have ultraviolet 
divergences. 
The validity of this assumption is not obvious, and it should be confirmed
by a concrete calculation. 

Because we carry out a linear analysis of the fluctuations, 
we replace $f(\lambda,D)$ by $f_0(\lambda)\equiv f(\lambda,D=0)$.
Then the problem is now reduced to deriving the form of the function 
$f_0(\lambda)$.  
However, because it is still difficult to find a general solution 
even in this
simplified problem, we study two specific cases, $\lambda \ll 1 $ and 
$\lambda \gg 1 $. 

When $\lambda \ll 1$, all quantities may be expanded in powers 
of $\lambda$.
Due to the reflection symmetry, we have 
$f_0(\lambda)  \simeq  c_0 \lambda^2 $.
Here $c_0$ is a constant whose value is calculated to be $1/1152\pi$ below.
We thus obtain 
\begin{equation}
p_y-p_y^{\rm eq}=c_0 TL \left( \frac{S}{\nu}\right)^2.
\label{py1}
\end{equation}
Note that this $L$ dependence, which implies the breakdown of the 
intensive nature, is compatible with the result by Aoki 
and Kusnezov \cite{Aoki}.
At the end of this Letter, we discuss that this anomalous behavior can be 
attributed to the long-range correlation of momentum fluctuations.

Next, we shall consider the opposite case, $\lambda \gg 1 $.  
Note that this condition does not necessarily imply the strong shear. 
For example, this asymptotic case is realized by taking the limit 
$L \to \infty$, while keeping the shear rate $S$ small, in which case 
the shear stress obviously remains small. 
Note also that planar Couette flow is linearly stable for all values of 
$\lambda$.
If we encounter no infrared divergence in the 
calculation of $p_y-p_y^{\rm eq}$,  we obtain 
$f_0(\lambda)=c_1 \lambda^{3/2}$, because $p_y-p_y^{\rm eq}$ should be  
independent of $L$. 
In this case we have
\begin{equation}
p_y-p_y^{\rm eq}=c_1 T \left( \frac{S}{\nu}\right)^{3/2}.
\label{py2}
\end{equation}
This $S$ dependence is the same as that obtained by Kawasaki and Gunton.
As we demonstrate below indeed, no infrared divergence is accompanied in 
the calculation of $p_y-p_y^{\rm eq}$, and that calculation yields 
(\ref{py2}) with $c_1 = 1.06 \times 10^{-2}$.

\paragraph*{Technical details :}

We first consider the case $\lambda \rightarrow \infty$ and begin by 
performing the Fourier expansion 
\begin{equation}
v_y(\vecr)=\int \frac{d\veck}{(2\pi)^3} e^{i \veck\vecr} 
\hat v_y(\veck).
\label{fex}
\end{equation}
In a statistically steady state, an equal-time correlation 
function can be defined as 
\begin{equation}
\bra \hat v_y(\veck) \hat v_y(\veck') \ket=
C_{yy}(\veck)\delta^3(\veck+\veck').
\end{equation}
Equation (\ref{pre}) is then rewritten in the form
\begin{equation}
p_y =p_{\rm B}+\rho 
\int \frac{d\veck}{(2\pi)^3}C_{yy}(\veck),
\label{pres}
\end{equation}
where we have used the result $\bra (\partial_l v_m )(\partial_m v_l )\ket=0$,
which is readily obtained by $\partial_lv_l = 0$ and (\ref{ni}).
From (\ref{pcon}) and (\ref{ns}), we derive the equation of the 
correlation function $C_{yy}(\veck)$:
\begin{eqnarray}
-Sk_x\pder{C_{yy}(\veck)}{k_y} &=& 
-2\left(\nu k^2-2S\frac{k_x k_y}{k^2}\right)C_{yy}(\veck) \nonumber \\
	&+& \frac{2\nu T}{\rho}(k_x^2+k_z^2).
\label{corr}
\end{eqnarray}
We can solve this equation by introducing the new wavevector 
$\veck'= \veck+Stk_x {\vece}_y$, where $t$ is a fictitious 
time parameter. 
Using the standard manipulation \cite{Dufty, OK}, we obtain
\begin{eqnarray}
C_{yy}(\veck)&=& \frac{2\nu T}{\rho}(k_x^2+k_z^2) 
\int_0^\infty dt e^{-2\nu k^2(t+t^2 S\hat k_x\hat k_y+t^3 S^2\hat k_x^2/3) }
\nonumber \\
&\phantom{=}&
(1+2tS \hat k_x\hat k_y+t^2 S^2\hat k_x^2)^2.
\label{col2}
\end{eqnarray}

Combining (\ref{pres}) and (\ref{col2}), we find an integral expression of 
the pressure \cite{other}.
When $S=0$, this merely gives the equilibrium pressure $p_y^{\rm eq}$,
which can be written as
\begin{equation}
p_y^{\rm eq}= p_{\rm B}+b T\ell^{-3}. 
\label{eqpy}
\end{equation}
Here $\ell$ is introduced as a short scale cut-off to avoid the ultraviolet 
divergence, and $b$ is a numerical constant. 
Because $p_{\rm B}$ remains undetermined, this ultraviolet divergence 
is renormalized into it appropriately so that $p_y^{\rm eq}$ coincides 
with the observable equilibrium pressure.
In a nonequilibrium case $(S > 0)$, we find that the correction to the 
equilibrium pressure is actually expressed in the form (13), 
and $c_1$ is given by
\begin{equation}
c_1=\sqrt{\frac{\pi}{32}}\frac{1}{(2\pi)^3} \int_0^\infty dt 
\frac{1}{t^{3/2}}g(t),
\label{goal}
\end{equation}
with
\begin{equation}
g(t)=\int d\Omega A(\Omega)\frac{B(\Omega)+tC(\Omega)}
{\left[1+tB(\Omega)+\frac{t^2}{3} C(\Omega)\right]^{3/2}}.
\label{fdef}
\end{equation}
Here $\Omega=(\theta,\phi)$, $d\Omega=d\phi d\theta \sin\theta$ and
$A$, $B$, $C$ are polynomials of $\sin\theta$, $\cos \phi$ and $\sin\phi$, 
which are determined from (\ref{col2}). 
From numerical integrations, we obtain $c_1=1.06\times 10^{-2}$. 

In the case $\lambda \ll 1$, we can utilize the calculation performed
above by simply replacing $\int dk_y (2\pi)^{-1}e^{i k_y y}\cdots$ with 
$\sum_n  L^{-1}e^{i 2\pi n  y/L} \cdots$ in (\ref{fex}). 
In this way, an expression similar to (\ref{col2}) with the replacement 
$k_y \to 2\pi/L$ is found.
Then the expansion of this correlation function in powers of $\lambda$
leads to the desired form of the nonequilibrium pressure 
in (\ref{py1}), whose numerical factor is obtained as 
$c_0=1/1152\pi \approx 2.8\times10^{-4}$.

\paragraph*{Long-range correlations :}

\begin{figure}[hbt]
  \begin{center}
  	\includegraphics[width=0.6455\linewidth]{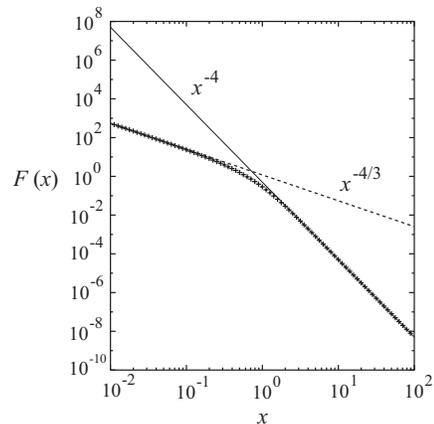}
  \end{center}
  \caption{Plots of numerically evaluated scaling function $F(x)$, where
  the correlation function $C_{yy}(k)$ is written in the form 
  $C_{yy}(k)=\rho^{-1}T(1+F(kl))$.
  Note that $\rho^{-1}T F(kl)$ corresponds to the nonequilibrium correction
  to the momentum correlation function.
  The solid line and the dotted line
  represent the asymptotic functions calculated from  
  the exact integral form of $F(x)$ for large $x \gg 1$ and for 
  small $x \ll1$, respectively. These asymptotic functions
  are equivalent to the results (\ref{col4}) and (\ref{col5}).}
  \label{fig:cyy}
\end{figure}

Here we demonstrate that the anomalous forms of the pressure 
(\ref{py1}) and (\ref{py2}) are closely related to the long-range correlations 
of momentum fluctuations. 
For simplicity, we restrict our attention to the fluctuations 
with $\veck=(k,0,0)$.
Two asymptotic forms of the correlation function are given as, 
\begin{equation}
C_{yy}(k,0,0)
\sim \frac{T}{\rho}\left(1+\frac{1}{2}\frac{S^2}{\nu^2 k^4}\right),
\label{col4}
\end{equation}
for $k^2  \gg S/\nu$, and 
\begin{equation}
C_{yy}(k,0,0)\sim \frac{T}{\rho} \left(\frac{2}{3}\right)^{1/3}
\Gamma\left(\frac{2}{3}\right)
\frac{S^{2/3}}{\nu^{2/3} k^{4/3}},
\label{col5}
\end{equation}
for $k^2  \ll S/\nu$. 
Comparing these expressions with (\ref{py1}) and (\ref{py2}), the
dependence of the pressure on the long-range fluctuations in each case
becomes clear. 
The $1/k^4$ dependence in (\ref{col4}) reflects highly anomalous 
behavior of the fluctuations with small wavenumbers \cite{Sen}. 
Note that there is a range of wavenumbers that satisfy 
$k^2  \gg S/\nu$ but are still much smaller than the 
characteristic wavenumber of equilibrium density fluctuations.
On the other hand, (\ref{col5}) shows that this long-range correlation 
is suppressed at scales larger than $l\equiv\sqrt{\nu/S}$, 
crossing over to a weaker correlation (See also Fig. \ref{fig:cyy}). 
This is equivalent to the stronger power law decay of the correlation
function than $1/r$ in a real space \cite{Dufty2}.
A new length scale $l$ characterizes this crossover, which is
intrinsic in the nonequilibrium system considering now.

These results provide the following physical picture. The momentum
fluctuations exhibit the long-range correlation described by (\ref{col4})
and (\ref{col5}). 
When $L \ll l$, this correlation yields the non-intensive 
contribution to the pressure given in (\ref{py1}). 
On the other hand, when $L$ is chosen to be sufficiently large, 
the long-range correlation is suppressed at scales larger than $l$,  
and this leads to the non-analytic shear rate dependence given 
by (\ref{py2}) instead. 

\paragraph*{Discussion:}
In a nonequilibrium system, an external field having a spatial
gradient (i.e. shear flow) induces the coupling of fluctuations with 
different wavevectors. 
In a perturbative expansion to the lowest order in shear rate, 
a correlation function has the same form as (\ref{col4}).
This may 
cause an infrared divergence in the calculation of $p_y-p_y^{\rm eq}$.
Roughly speaking, (\ref{py1}) is obtained when the cut-off scale is chosen as 
$L$, while (\ref{py2}) is obtained when it is chosen as $l$.
We have demonstrated the validity of this intuitive 
argument by computing the correlation function rigorously.

The predicted finite-size dependence of the pressure (\ref{py1})
under the condition $\lambda \ll 1$ is fairly striking.
Since $l=\sqrt{\nu/S}$ may be less than 1 mm for water 
at standard temperature and pressure and for experimentally accessible 
shear rate, it is possible to design an experimental device 
corresponding to the the condition $\lambda \ll 1$.
On the other hand, from (\ref{eqd}), (\ref{py2}) and (\ref{eqpy}),  
$(p_y-p_y^{\rm eq})/p_y^{\rm eq}$ is found to be proportional to 
$\epsilon^{3/2}$ in the case $\lambda \gg 1$, and $\epsilon^2$ 
in the case $\lambda \ll 1$, respectively.
This indicates that the nonequilibrium correction vanishes 
in local equilibrium states (i.e., $\epsilon \to 0$), as expected. 
For example, $\epsilon$ is less than $10^{-8}$ when the shear rate 
is $10^3 \,{\rm sec}^{-1}$ for water considered above.
Thus, unfortunately, the nonequilibrium correction may be 
too small to observe for simple fluids under ordinary experimental 
conditions in either case.

This fact, however, does not eliminate the fundamental 
significance of the present study. 
In constructing a theoretical framework of non-equilibrium statistical 
mechanics, 
this '$\epsilon$-effect' must be taken into consideration, because a 
statistical distribution can be reduced to a local canonical 
ensemble only in the limiting case $\epsilon \to 0$.
Furthermore, making use of the experimental technique 
developed in the micro- and  nano-fluid studies may enable us to detect 
the novel nonequilibrium effect for real fluids in the near future.
We also expect that our findings stimulates further numerical studies 
such as molecular dynamic simulations. 

Recently, 
Marcelli et al. reported the analytic dependence of pressure in shear flow 
observed in simulations of non-equilibrium molecular dynamics \cite{Mar}. 
Because the shear conditions were chosen to satisfy 
$\epsilon \ge 1 $ and $\lambda \gg1 $ in their simulations, 
their result is not directly comparable to ours. 
However, we believe that the qualitative feature of our result does not change 
even in the case $\epsilon \ge 1 $. 
We suspect that their model system does not exhibit long-range
momentum correlation because of the absence of the local 
conservation of momentum. That might be the reason why 
their result is inconsistent with the existing mode-coupling 
theories \cite{KG, YK, Ernst, Lutsko} and ours. 
We expect that measuring momentum correlation functions will help to
resolve the discrepancy, probably favorably with the mode-coupling theory.

Finally, we give a general remark on the possibility of describing an equation 
of state for a non-equilibrium steady state using a thermodynamic function.
An experimental test of this possibility has recently been proposed in \cite{ST}, 
in which the intensivity of the pressure is postulated.
Therefore, the recovery of the intensivity in our calculation has a significant
meaning with regard to the construction of thermodynamics extended to steady 
states. 

The authors acknowledge K. Kawasaki for stimulating conversations.
This work was supported by grants from the Ministry of Education,
Science, Sports and Culture of Japan, No. 14654064.


\begin{thebibliography}{10}
\bibitem{other}

Other components of the pressure as well as the dynamic shear viscosity 
can be calculated in the same way.
\end{thebibliography}
\end{document}